
\input phyzzx
\font\smallrm=cmr8
\input epsf


\REF\SGP {R. Sorkin, Phys. Rev. Lett. {\bf 51} (1983) 87 ; D. Gross and M.
Perry,
Nucl. Phys. {\bf B226} (1983) 29.}
\REF\GGM {G.W. Gibbons, Nucl. Phys. {\bf B207} (1982) 337;
G.W. Gibbons and K. Maeda, Nucl. Phys. {\bf B298} (1988) 741.}
\REF\GHS{ D. Garfinkle, G. Horowitz and A. Strominger, Phys. Rev. {\bf D43}
(1991) 3140; {\bf D45} (1992) 3888(E).}
\REF\GH {G.W. Gibbons and C.M. Hull, Phys. Lett. {\bf 109B} (1982) 190.}
\REF\HS {G.T. Horowitz and A. Strominger, Nucl. Phys. {\bf B360} (1991) 197.}
\REF\DS {M.J. Duff and K.S. Stelle, Phys. Lett. {\bf 253B} (1991) 113.}
\REF\DGHR{ A. Dabholkar, G.W. Gibbons, J.A. Harvey and F. Ruiz-Ruiz, Nucl.
Phys. {\bf
B340} (1990) 33.}
\REF\DGT {M.J. Duff, G.W. Gibbons and P.K. Townsend, Phys. Lett.
{\bf 332 B} (1994) 321.}
\REF\CVETIC{ M. Cvetic, S. Griffies and S.J. Rey, Nucl. Phys. {\bf B381} (1992)
302; M. Cvetic and S. Griffies, Phys. Lett. {\bf B285} (1992) 27; M. Cvetic,
Phys.
Rev. Lett. {\bf 71} (1993) 815.}
\REF\HH {J.H. Horne and G.T Horowitz, Nucl. Phys. {\bf B368} (1992) 444.}
\REF\GT {G.W. Gibbons and P.K. Townsend, Phys. Rev. Lett. {\bf 71} (1993)
3754.}
\REF\Gu {R. G{\" u}ven, Phys. Lett. {\bf 276B} (1992) 49.}
\REF\vB {P. van Baal, F.A. Bais and P. van Nieuwenhuizen, Nucl. Phys. {\bf
B233} (1984) 477.}
\REF\GKLTT {G.W. Gibbons, D. Kastor, L. London, P.K. Townsend and J. Traschen,
Nucl.
Phys. {\bf B416} (1994) 850.}
\REF\GL {R. Gregory and R. Laflamme, Phys. Rev. Lett. {\bf 70} (1993) 2837;
``The instability of charged black strings and $p$-branes", DAMTP-R-94-7,
hep-th/9404071.}
\REF\BI { U. Bleyer and V.D. Ivashchuk, Phys. Lett. {\bf 322B} (1994) 292.}
\REF\DL {M.J. Duff and X. Lu, Nucl. Phys. {\bf B 411} (1994) 301.}
\REF \BHKT { D. Brill, G. Horowitz, D. Kastor, and J. Traschen,
Phys. Rev. {\bf D49} (1994) 840.}
\REF\Wi {E. Witten, Commun. Math. Phys. {\bf 80} (1981) 381.}
\REF\BKK{L. Bombelli, R. Koul, G. Kunstatter, J. Lee and R. Sorkin,
Nucl. Phys. {\bf B289} (1987) 735.}
\REF\TvN {P.K. Townsend and P. van Nieuwenhuizen, Phys. Lett. {\bf 125B} (1983)
41.}


\Pubnum={R/94/28\cr UCSBTH-94-35\cr hep-th/9410073}
\pubtype{}
\date{October, 1994}

\titlepage

\title{Higher-dimensional resolution of dilatonic black hole singularities}

\author{G. W. Gibbons\foot{Permanent address: DAMTP, University of Cambridge,
Silver St., Cambridge, U.K.} and Gary T. Horowitz\foot
{Current address:
Physics Department, University of California, Santa Barbara, CA. 93106;
email: gary@cosmic.physics.ucsb.edu} }
\address{Isaac Newton Institute, Cambridge, U.K.}
\andauthor{P. K. Townsend}
\address{DAMTP, University of Cambridge, U.K.}

\vskip 1cm

\abstract {We show that the four-dimensional extreme dilaton black
hole with dilaton coupling constant $a= \sqrt{p/(p+2)}$
can be interpreted as a {\it completely non-singular}, non-dilatonic,
black $p$-brane in $(4+p)$ dimensions provided that $p$ is {\it odd}.
Similar results are obtained for multi-black holes and dilatonic extended
objects in higher spacetime dimensions. The non-singular
black $p$-brane solutions include the self-dual three brane of
ten-dimensional N=2B supergravity and a multi-fivebrane solution of
eleven-dimensional supergravity. In the case of a supersymmetric
non-dilatonic $p$-brane solution of a supergravity theory, we show that
it saturates a bound on the energy per unit $p$-volume.
\endpage

\pagenumber=2



\def\om{\omega}



\chapter{ Introduction}

Given certain mild conditions on the matter stress tensor, the singularity
theorems
of Penrose and Hawking guarantee the existence of spacetime singularities in a
broad
class of four-dimensional solutions of Einstein's equations describing
gravitational collapse. The occurrence of such
inevitable singularities is a good indication that `new physics' is required.
It
is commonly believed that the new physics in question involves quantum
mechanics, so that a proper understanding of how spacetime singularities are
resolved must await a consistent theory of quantum gravity. While this may well
be
true, it is also possible that new {\it classical} physics intervenes,
at some energy
scale below  the Planck mass, in such a way that singularities are resolved.
To remove all singularities, one needs a fundamental change in our
description of classical gravity, such as perhaps the one provided by
string theory.
However certain singularities  can
be resolved by simply passing to a
higher-dimensional theory of gravity for which spacetime is only
{\it effectively} four-dimensional below some compactification scale.

For example, consider any  nonsingular
five dimensional solution in Kaluza-Klein theory,
which has a spacelike isometry  with a fixed point.
Then its dimensional reduction along the isometry yields a four dimensional
solution with a curvature singularity. If one did not know the origin of
this solution, it would not be obvious that the singularity was simply
the result of trying to project a regular five dimensional solution into four
dimensions. The best known example of this phenomenon
is the Kaluza-Klein monopole [\SGP], in which one starts with the product of
time and the Euclidean Taub-NUT solution. But one can construct  other
examples by e.g. starting
with the product of time and the Euclidean Schwarzschild solution, or even
starting with five dimensional Minkowski spacetime and reducing along a
rotation (although in this last case,
the resulting four-metric is not asymptotically
flat). In all of these examples,
the four dimensional singularity is timelike.

One can also remove certain spacelike singularities this way.  The simplest
example is again to start with five dimensional Minkowski spacetime
and reduce along the orbits of a boost (considering only the region where
the boost is spacelike). The resulting four dimensional
metric describes a spatially flat Robertson-Walker universe with an initial
or final singularity.
Notice that in this case, if one
identifies points to make the orbit of the boost compact (and the spacetime
appear four dimensional), the  five
dimensional space  has a conical singularity at the origin and is no longer
geodesically complete.  So the curvature singularity is
not completely resolved, but replaced by a much milder conical singularity.

We will show that the curvature singularity
in many of the extreme dilaton black holes
[\GGM, \GHS] can be resolved in this way. These black holes have a null
singularity which becomes a regular horizon in the higher dimensional
spacetime. More precisely, when the
dilaton coupling constant $a$ (defined below in (2.1)) takes one of the special
values
$$
a =\sqrt{{p\over p+2}}
\eqn\onea
$$
for integer $p$,  the extreme dilaton black hole can be
re-interpreted as a black $p$-brane in $(4+p)$ dimensions. When $p$ is
even, the $p$-brane resembles the extreme Reissner-Nordstr\"om solution
(which is included as the special case $p=0$) in that there is
still a curvature singularity
inside the horizon. However, for $p$ odd, the spacetime behind the
horizon is isometric to the region outside, and the solution is completely
nonsingular.\foot{These solutions are analogous to the case
of the spacelike singularity
above: if one compactifies the extra $p$ dimensions, the spacetime becomes
geodesically incomplete even though the curvature remains bounded everywhere.}
For example, if
$p=1$ we have $a=1/\sqrt{3}$, and the four-dimensional extreme
dilaton black hole can be
interpreted as the double-dimensional reduction of a non-singular
five-dimensional black string. This is called `double'-dimensional
reduction, in contrast
to the simple dimensional reduction of the Kaluza-Klein monopole example,
because in reducing
the spacetime dimension from five to four we simultaneously reduce the
dimension of
the extended object from one to zero. It is precisely the reverse procedure
that
allows a resolution of the spacetime singularities of the dilaton black hole
for the
above values of the dilaton coupling constant. The special case of
$a=1/\sqrt{3}$ is
of particular interest because, like the Kaluza-Klein monopole for
$a=\sqrt{3}$, the
five-dimensional black string is also a solution of the pure
five-dimensional supergravity.
In this case one can derive a bound on the energy per unit length,
analogous to that of
[\GH] for particle-like solutions of four-dimensional Maxwell-Einstein theory.
This
bound is saturated by the black string solution. In contrast the same methods
fail to establish a similar bound for the black  $p$-branes in $4+p$
dimensions if $p>1$, as might have been expected from the absence of an
appropriate
supergravity theory underlying these cases.

The black $p$-branes we will construct differ from  those in [\HS]
in that we consider solutions
to a theory without
a dilaton in the higher dimensional space. The four dimensional
dilaton arises from the dimensional reduction.
Some examples of non-dilatonic $p$-branes have been discussed previously.
One is the membrane solution of eleven-dimensional supergravity [\DS],
whose double-dimensional reduction
yields the fundamental string solution [\DGHR] of
ten-dimensional supergravity. This is analogous to the case of \onea\ with
$p$ even. The singularity in the fundamental string becomes a regular
horizon in higher dimensions, but there is another curvature singularity
inside [\DGT].

The existence of solutions with horizons but no singularities is certainly
surprising, but not without precedent. For example, domain wall
solutions of $N=1$ supergravity have been found of this type
[\CVETIC]. Of more relevance here is the metric representing an extreme black
string solution in three dimensions
that is part of an exact conformal field theory
[\HH]. This is very similar to the five-dimensional string metric that resolves
the
singularities of the $a=1/\sqrt{3}$ dilaton black hole. In fact, if
one discards the angular part of the five-metric
the resulting three-metric approaches
the metric of [\HH] near the horizon. This metric was shown in [\HH] to be
non-singular, although good coordinates for the region containing the horizon
were
not given. Here we present a similar analysis for the
$p$-brane solutions under discussion, including details of how one finds good
coordinates near the horizon.

The analysis just described is readily generalized to extreme,
dilaton black holes, or extended objects, in a $d$-dimensional spacetime.
One again finds that in certain cases, the singularity in these solutions
can be removed via re-interpretation as a higher-dimensional object in
a higher-dimensional spacetime. The mechanism by which this is achieved is best
understood by considering the asymptotic metric and dilaton near the singular
hypersurface [\GT, \DGT]. The asymptotic metric is conformal to
the product of a sphere with a
lower-dimensional anti-de Sitter spacetime. The dilaton is singular at a
Killing horizon of the anti-de
Sitter space, but in such a way that it can be viewed as the logarithm of
additional diagonal components of a higher-dimensional adS-metric. This
involves
the re-interpretation of the solution as a $p$-brane in a higher dimensional
theory  without a dilaton and the singularity of the dilaton at the horizon is
then equivalent to the vanishing of these additional metric components there.
This
is a mere coordinate singularity and the metric can be analytically continued
through the horizon to an interior region. One can also start
with non-extreme black holes and construct new non-extreme black $p$-brane
solutions in higher dimensions as we will show. However, in this case
the singularity inside the black hole horizon is not resolved by lifting
to higher dimensions.

A further generalization to multi-$p$-branes is also possible. We find that
the singularities of four-dimensional multi-dilaton black hole solutions are
again
resolved by their
interpretation as a multi-$p$-brane solution in $4+p$ dimensions if $p$
is odd. For $d>4$, the situation is more subtle. In some cases, the singularity
is completely resolved, while in others the multi-$p$-brane
has singularities inside
the horizon even though the single $p$-brane was completely nonsingular.
Finally, there are cases in which the horizons of the multi-$p$-brane
seem to have only finite differentiability. One case which is completely
nonsingular is
a multi-version of G{\" u}ven's eleven-dimensional fivebrane [\Gu], and we
show how this solution nicely realises the idea of
`local compactification' [\vB].

A question of obvious interest is whether the $p$-brane solutions found here
are
stable. A strong indication of stability would be a generalization of
the Bogomol'nyi-type bounds [\GH,\GKLTT] established for black holes. We note
that
while extreme black holes and $p$-brane solutions in arbitrary dimension are
generally characterised by a simple Bogomolnyi-type relation between their mass
and
charge (per unit $p$-volume in the $p$-brane case) there is no evidence in
general
that there is a corresponding Bogomolnyi-type {\it bound} which applies to all
solutions. One expects to be able to
derive such a bound in the case of a solution of a field theory that is the
bosonic
sector of a supergravity theory, because in this case the bound is essentially
implied by the algebra of supersymmetry. Indeed, we shall show that the methods
of
[\GH, \GKLTT] yield the expected bound precisely in these circumstances, {\it
and in
no others}. This strongly suggests that supersymmetric extended object
solutions of
supergravity theories will not suffer from the type of instability recently
found
[\GL] to afflict certain non-extreme extended object solutions of
higher-dimensional
gravity. It leaves open the interesting question of the stability of
non-supersymmetric extreme black holes or $p$-branes.


\chapter{Resolution of extreme dilaton black hole singularities}

The four-dimensional action for the spacetime metric $g$, one-form abelian
gauge
potential $A$, with two-form field-strength $F_2$, and dilaton
$\phi$ with dilaton coupling constant $a$, is
$$
S= {1\over 16\pi G}\int\! d^4x\, \sqrt{-g}\big\{ R - 2(\nabla\phi)^2
- e^{-2a\phi} F_2^2\big\}\ .
\eqn \twoa
$$
We may suppose without loss of generality that $a\ge 0$ since the field
redefinition
$\phi\rightarrow -\phi$ effectively changes the sign of $a$. The Euler-Lagrange
(E-L) equations of \twoa\ admit a magnetically-charged black hole solution
[\GGM,\GHS] whose extremal limit is
$$
\eqalign{
ds^2 &= -\Big (1-{\mu\over r}\Big )^{2\over 1+a^2} dt^2  + \Big(1-{\mu\over
r}\Big)^{-{2\over 1+a^2}} dr^2  + \Big(1-{\mu\over r}\Big)^{2a^2\over 1+a^2}
r^2
d\Omega_2^2 \cr
e^{a\phi} &= \Big(1-{\mu\over r}\Big)^{-{a^2\over 1+ a^2}} \cr
F_2 &= Q\varepsilon_2 \ ,}
\eqn\twob
$$
where $d\Omega_2^2$ is the standard metric on the unit two-sphere and
$\varepsilon_2$ is its volume 2-form. The constant $\mu$ is related to the
ADM mass M, and magnetic charge $Q$, in units for which $G=1$, by
$$
\mu = \sqrt{1+ a^2}\ |Q|= (1+ a^2)\, M\ .
\eqn\atwob
$$
This solution  saturates the bound [\GKLTT]
$$
M\ge {1\over \sqrt{1+a^2}}|Q|
\eqn\twoc
$$
on all asymtotically flat solutions that can be formed from gravitational
collapse.
It is also singular at $r=\mu$. We shall now show how this singularity can be
resolved for the values of $a$ given in \onea\
by re-interpreting the solution as representing
an extended object in a higher-dimensional spacetime. The singularity at
$r=\mu$
becomes simply a coordinate singularity at a degenerate Killing horizon.
Furthermore, we shall show that when $p$ is odd,
the maximal analytic extension of this higher-dimensional metric is
completely non-singular.

Consider the $(4+p)$-dimensional Einstein-Maxwell action
$$
S_{(4+p)} = \int\! d^{4+p}x\sqrt{-g}\big\{ R - F_2^2\big\}\ .
\eqn\twod
$$
We assume that the $(4+p)$-metric and Maxwell field take the form
$$
\eqalign{
ds^2_{(4+p)} &= e^{2\alpha\phi(x)}d{\bf y}\cdot d{\bf y} \ +\ e^{2\beta\phi(x)}
g_{\mu\nu}(x) dx^\mu dx^\nu \cr
F^{(4+p)}_2 &= {1\over2} F_{\mu\nu}(x) dx^\mu \wedge dx^\nu\ ,}
\eqn\twoe
$$
where $x^\mu$ are the coordinates of a four-dimensional submanifold and ${\bf
y}$ are
the $p$ additional coordinates. Notice that both the metric and Maxwell field
are independent of ${\bf y}$, and we do not include any $x^\mu,
{\bf y}$ cross terms in the metric
which would give rise to additional gauge fields in
four dimensions. With this form of the metric and Maxwell field, the action
\twod\ reduces to the following  four-dimensional action
$$
\eqalign{
S= {1\over 16\pi G}\int\! d^4x\, &\sqrt{-g}e^{(2\beta + p\alpha)\phi}
\big\{ R - 2(p\alpha + 3\beta)\nabla^2\phi\cr
&-(6\beta^2 + p\alpha^2 + p^2\alpha^2 + 4p\alpha\beta)(\nabla\phi)^2 -
e^{-2\beta\phi}F_2^2\big\}\ ,}
\eqn\twof
$$
In order that the four-metric have the
canonical, Einstein-Hilbert, action, we choose $\alpha$ and $\beta$ such that
$$
2\beta + p\alpha =0\ .
\eqn\twog
$$
In this case the $\nabla^2\phi$ term is a surface term which, for present
purposes,
we may neglect. Hence
$$
S= {1\over 16\pi G}\int\! d^4x\, \sqrt{-g}
\big\{ R -{p(p+2)\over2}\alpha^2(\nabla\phi)^2 - e^{p\alpha\phi}F_2^2\big\}\ .
\eqn\twoh
$$
By choosing
$$
\alpha = -{2\over \sqrt{p(p+2)}}
\eqn\twoi
$$
we recover the action \twoa\ with dilaton coupling constant
$$
a= \beta= \sqrt{p\over p+2}\qquad p=0,1,2,3,\dots
\eqn\twoj
$$
For these values of $a$ {\it any} solution $g_{\mu\nu}(x)$, $\phi(x)$ and
$F_2(x)$
of the E-L equations of \twoa\ can be interpreted as a solution of the E-L
equations of the $(4+p)$-dimensional action \twod\ with metric and two form
$$
\eqalign{
ds^2 &= e^{2(a-a^{-1})\phi(x)} d{\bf y}\cdot d{\bf y} \ +\
e^{2a\phi(x)}g_{\mu\nu}(x)dx^\mu dx^\nu \cr
F_2 &= {1\over2}F_{\mu\nu}(x)dx^\mu\wedge dx^\nu \ .}
\eqn\twok
$$

For example, the magnetically-charged extreme dilaton black hole solution
\twob\ can be viewed as the double-dimensional reduction of
$$
ds^2 = \Big(1-{\mu\over r}\Big)^{2\over p+1}(-dt^2 + d{\bf y}\cdot d{\bf y})
+\Big(1-{\mu\over r}\Big)^{-2}dr^2 + r^2d\Omega_2^2
\eqn\twol
$$
which is the metric of a black $p$-brane aligned with the
$y$-axes. The singularity of the $(4+p)$-metric at $r=\mu$ is now merely a
coordinate singularity. To see this, it is useful to first determine the
behaviour of the metric in the neighbourhood of $r=\mu$ by introducing a new
radial coordinate $\omega$ by the relation
$$
(p+1) \mu\omega = \Big(1-{\mu\over r}\Big)^{1\over p+1}
\Leftrightarrow r= {\mu\over 1-[(p+1)\mu\omega]^{p+1}}
\eqn\twom
$$
which is valid for $r > \mu$, to arrive at
$$
\eqalign{
ds^2 &= (p+1)^2 \mu^2\Big[ \omega^2(-dt^2 + d{\bf y}\cdot d{\bf y}) +
\big(1-[(p+1)\mu\omega]^{p+1}\big)^{-4}\Big({d\omega\over \omega}\Big)^2
\Big]\cr
&\qquad +  \big(1-[(p+1)\mu\omega]^{p+1}\big)^{-2}\mu^2 d\Omega_2^2 \ .\cr}
\eqn\twon
$$
As $\omega\rightarrow 0$ we have
$$
ds^2 \sim (p+1)^2 \mu^2\Big[ \omega^2(-dt^2 + d{\bf y}\cdot d{\bf y}) +
\Big({d\omega\over \omega}\Big)^2 \Big] + \mu^2 d\Omega_2^2\ .
\eqn\twona
$$
This asymptotic metric is the product of $S^2$ with $(p+2)$-dimensional anti-de
Sitter space, $(adS)_{(p+2)}$, as we now show.

Anti-de Sitter space $(adS)_{(p+2)}$ can be viewed as the hypersurface
$$
X_0^2 - {\bf Y}\cdot {\bf Y} + X_+ X_- = 1
\eqn\twoo
$$
in a $(p+3)$-dimensional flat spacetime with coordinates $(X_0, X_\pm =
X_1 \pm X_2, {\bf Y})$ and  metric
$$
ds^2 = -dX_0^2 +d{\bf Y}\cdot d{\bf Y} - dX_- dX_+
\eqn\twop
$$
of signature $(p+1,2)$. If we introduce hypersurface coordinates
(t,{\bf y},$\omega$) defined
by
$$
\eqalign{
\omega &= X_-\cr
t\omega &= X_0\cr
{\bf y}\omega &= {\bf Y}\ ,}
\eqn\twoq
$$
then
$$
X_+ = {1+\omega^2(|{\bf y}|^2-t^2)\over\omega}\ ,
\eqn\twor
$$
and the induced hypersurface metric is found to be
$$
ds^2 = -\omega^2(dt^2 -d{\bf y}\cdot d{\bf y}) + \omega^{-2} d\omega^2
\eqn\twos
$$
which is the metric obtained previously. This form of the
$(adS)_{(p+2)}$ metric is analogous to the spatially flat slicing of
de Sitter space. Since the coordinates $(X_0, X_\pm, {\bf Y})$ evaluated
on the hypersurface are necessarily smooth,
it is clear from \twoq\ that $\omega$ remains a good coordinate at the
horizon $\om =0$, while $t$ and ${\bf y}$ become ill defined. It is also
clear that the symmetry generating translations of {\bf y} has a fixed point
at the horizon, so that if we make these coordinates periodic,
the spacetime is no longer geodesically complete.

We now observe that since $r$ is an analytic function of $\omega$ near $\om=0$,
the metric
can be extended analytically through the horizon at $\omega=0$ and hence the
$p$-brane metric \twol\ is completely regular at $r=\mu$. Furthermore, since
$r$ is a function of $\om^{p+1}$,  if
$p$ is {\it odd} then the full metric
is invariant under the reflection
$$
\omega \rightarrow -\omega\ .
\eqn\twot
$$
This implies that the analytic extension through $\omega=0$ leads to an
interior
region that is isometric to the exterior region, which is singularity free.
This
interior region has its own horizon through which we can continue the analytic
extension as before. Proceeding in this way one obtains a completely
non-singular
maximal analytic extension of the $p$-brane metric of \twol. Thus, as claimed,
{\it
all} singularities of the four-dimensional dilaton black hole are resolved
in this way whenever the dilaton coupling constant $a$ takes one of the values
$$
a= \sqrt{p\over p+2}\qquad p=1,3,5,\dots\ .
\eqn\twou
$$
The global structure of these solutions is illustrated in the
Carter-Penrose diagram of Figure 1, below.
\midinsert
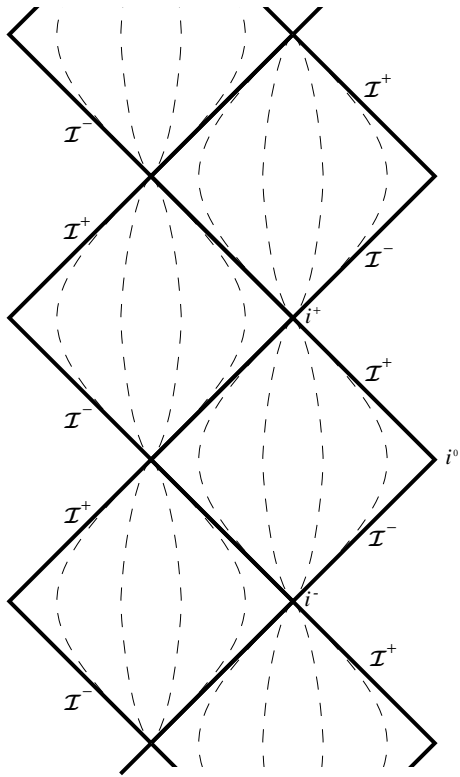
\epsffile{CPD}\centerline{\smallrm Fig. 1. CP diagram for the maximal analytic
extension of the metric of (2.13) for odd p.}
\centerline{\smallrm Each point on the diagram represents the product of a
2-sphere with a p-plane.}
\centerline{\smallrm The dotted lines are orbits of the timelike Killing vector
field ${\scriptstyle \partial/\partial t}$.}
\endinsert

If $p$ is even rather than odd, the singularity of the four-dimensional metric
at $\omega=0$, i.e. $r=\mu$, is still resolved by the higher-dimensional
interpretation, but on passing through the horizon one finds an interior region
with
a curvature singularity. The simplest example of this is the trivial case
$p=0$, in
which case the `higher-dimensional' spacetime is in fact also four-dimensional
and
the `higher-dimensional' metric is just the extreme Reissner-Nordstr\"om
metric.

It is amusing to note that the value $a=1$, which
is of interest in string theory,
corresponds to the limit $p\rightarrow \infty$. Thus in some sense, the
extreme black hole in string theory is the reduction of an infinite dimensional
extended object (see also [\BI]).

A generalization of the above results is possible by allowing the
two-form
$F_2$ in four dimensions to be the appropriate projection of a $(2+m)$-form
$F_{2+m}$ ($m\le p$) in the higher dimensional spacetime,
but simple results are obtained only when $m=0$ or $m=p$.
The former case is the one just analysed. In the latter case, the
$(4+p)$-dimensional action is
$$
S_{(4+p)}  = \int\! d^{4+p}x \sqrt{-g}\{ R - {2\over (p+2)!} F_{p+2}^2\}\ .
\eqn\twov
$$
For a $(4+p)$-metric of the form given in \twoe\ and a $(p+2)$-form with Hodge
dual
$$
{}^\star F_{p+2} = {1\over2}\tilde F_{\mu\nu}(x) dx^\mu\wedge dx^\nu\ ,
\eqn\twow
$$
the $(4+p)$-dimensional action again reduces to the four-dimensional action
\twoa\
with dilaton coupling constant given by \onea\ if $2\beta +p\alpha=0$, as
before,
and if $\alpha$ is now taken to have the opposite sign:
$$
\alpha = + {2\over \sqrt{p(p+2)}}\ .
\eqn\twox
$$
It follows that electrically-charged extreme dilaton black holes with
$a=\sqrt{p/(p+2)}$ can
also be interpreted in $(4+p)$-dimensions as a $p$-brane solution  whose
metric is still given by  \twol\ but
instead of the two-form $F_2 = Q \epsilon_2$, one has its dual, which is
a $(p+2)$-form.
This electrically-charged $p$-brane
solution is still
non-singular on its Killing horizon and completely non-singular if
$p$ is odd.


\chapter{ Generalization to higher dimensions}

Given a solution of the $(d+p)$-dimensional Einstein equations with a horizon
and $p$-fold
translational symmetry, i.e. a black
$p$-brane, one can always find a dilatonic black hole
solution of $d$-dimensional gravity by double-dimensional reduction. As we have
seen from the preceding examples the resulting dilatonic black hole may have a
singularity where the non-dilatonic $p$-brane had a horizon. If the maximal
analytic
extension of the $p$-brane solution is non-singular it can be considered to
resolve
the singularities of the black hole. The examples discussed in the
previous section constitute the
special case for which $d=4$. Now we shall consider the general case. Note that
given a non-singular $(d+p)$-dimensional $p$-brane there is no necessity to
double-dimensionally reduce it by $p$ dimensions; one might equally wish to
reduce
it by $q\le p$ dimensions to arrive at a singular dilatonic $(p-q)$-brane in
$(d+p-q)$ dimensions. The singularities of these dilatonic extended objects are
equally  resolved by the
$(d+p)$-dimensional $p$-brane. Since the choice of $q$ is immaterial to the
main
result we shall discuss here only the case $q=p$, i.e.
$d$-dimensional dilatonic black holes.

We start from the $(d+p)$-dimensional action
$$
S = \int\! d^{(d+p)}x \sqrt{-g}\big\{ R - {2\over
(d-2)!}F_{d-2}^2\big\}
\eqn\threea
$$
As before we perform the reduction by $p$ dimensions by taking the
$(d+p)$-metric
and $(d-2)$-form $F_{(d-2)}$ to be
$$
\eqalign{
ds^2_{(d+p)} &= e^{2\alpha\phi(x)}d{\bf y}\cdot d{\bf y} \ +\ e^{2\beta\phi(x)}
g_{\mu\nu}(x) dx^\mu dx^\nu \cr
F_{(d-2)} &= {1\over (d-2)!} F_{\mu_1\cdots \mu_{d-2}}dx^{\mu_1}\cdots
dx^{\mu_{d-2}} \ ,}
\eqn\threeb
$$
where $x^\mu$ are the coordinates of the $d$-dimensional spacetime. Using the
formula
$$
\tilde R =\Omega^{-2}\big[ R - 2(d-1)\nabla^2 \ln \Omega - (d-1)(d-2)(\nabla
\ln\Omega)^2 \big]
\eqn\threec
$$
for two $d$-dimensional
metrics $g$ and $\tilde g$ related by $\tilde g =\Omega^2 g$, one readily
finds that the $d$-dimensional action is of the canonical Einstein-Hilbert form
provided that $\alpha$ and $\beta$ satisfy
$$
p\alpha + (d-2)\beta =0\ ,
\eqn\threed
$$
which generalizes \twog. We may then fix the normalization of the dilaton field
kinetic term by choosing $\alpha$ such that
$$
\alpha^2 = {2(d-2)\over p(d+p-2)} \ .
\eqn\threee
$$
The $d$-dimensional action is now
$$
S = \int\! d^dx \sqrt{-g}\big\{ R - 2(\nabla\phi)^2 -
{2\over (d-2)!}e^{-2a\phi} F_{d-2}^2\big\}\ ,
\eqn\threef
$$
where
$$
a = {(d-3)\sqrt{2p}\over \sqrt{(d-2)(d+p-2)}}\ .
\eqn\threeg
$$
The magnetically charged black hole solutions of the E-L equations of this
action are
[\HS]
$$
\eqalign{
ds^2_d &= -\Big[1-\Big({r_+\over r}\Big)^{d-3}\Big]\Big[1-\Big({r_-\over
r}\Big)^{d-3}\Big]^{1-(d-3)\gamma} dt^2 \cr
&+ \Big[1-\Big({r_+\over r}\Big)^{d-3}\Big]^{-1}\Big[1-\Big({r_-\over
r}\Big)^{d-3}\Big]^{\gamma -1} dr^2 + r^2\Big[1-\Big({r_-\over
r}\Big)^{d-3}\Big]^\gamma d\Omega_{(d-2)}^2 \cr
e^{a\phi} &= \Big[1-\Big({r_-\over r}\Big)^{d-3}\Big]^{-{(d-3)\gamma\over2}}\cr
F_{(d-2)} &= Q\varepsilon_{d-2} ,}
\eqn\threeh
$$
where  $\varepsilon_{d-2}$ is the volume form on the unit $(d-2)$-sphere,
$$
\gamma \equiv {2p\over (d-2)(p+1)}\ ,
\eqn\threei
$$
and the charge $Q$ is related to $r_\pm$ by
$$
Q^2 = {(d+p-2)(d-3) \over 2(p+1)}\  (r_+ r_-)^{d-3} \ .
\eqn\tthreei
$$
The $(d+p)$-metric is therefore
$$
\eqalign{
ds^2 &=- \Big[1-\Big({r_+\over r}\Big)^{d-3}\Big]
\Big[1-\Big({r_-\over r}\Big)^{d-3}\Big]^{1-p\over 1+p}dt^2 +
\Big[1-\Big({r_-\over r}\Big)^{d-3}\Big]^{2\over p+1}d{\bf y}\cdot d{\bf y}\cr
&+ \Big[1-\Big({r_+\over r}\Big)^{d-3}\Big]^{-1}
\Big[1-\Big({r_-\over r}\Big)^{d-3}\Big]^{-1} dr^2 + r^2d\Omega_{(d-2)}^2 }
\eqn\threej
$$
This is a new class of black $p$-brane solutions of the action \threea.
In the general nonextremal ($r_+ > r_-$) case, both the black holes \threeh\
and the black $p$-branes \threej\ have event
horizons at $r=r_+$ and curvature singularities at $r=r_-$.

However, the extremal limit $r_+ = r_- \equiv \mu$ is different.
In this case, the black hole horizon becomes singular, but the black
$p$-brane solution takes the form\foot{Correcting a typographical error of
[\DGT].}
$$
\eqalign{
ds^2 &= \Big[1-\Big({\mu\over r}\Big)^{d-3}\Big]^{2\over p+1}(-dt^2 + d{\bf
y}\cdot d{\bf y}) + \Big[1-\Big({\mu\over r}\Big)^{d-3}\Big]^{-2}dr^2 +
r^2d\Omega_{(d-2)}^2 \cr
F &= \sqrt{{(d+p-2)(d-3)\over 2(p+1)}}\ \mu^{d-3}\varepsilon_{d-2} }
\eqn\threel
$$
This metric is invariant under the full $(p+1)$-Poincar{\'e} group, which
is also true for the extremal limit of dilatonic black $p$-branes [\HS].
The only difference between \threel\ and
the metric of \twol\ is the power of $r$ and the
fact that the last term in \threel\ involves the metric
on the unit $(d-2)$-sphere in place of the 2-sphere. Since the horizon is
at a nonzero value of $r$, the different powers of $r$ is unimportant and
the previous analysis of the
global structure carries over to this case: the metric is non-singular at
the Killing horizon $r=\mu$, and good coordinates can be
found as before. The analytic continuation through this horizon is symmetric if
$p$
is odd and in this case the maximal analytic extension is completely
non-singular.
Near the horizon the metric is asymptotic to $(adS)_{(p+2)}\times S^{(d-2)}$.
The
Carter-Penrose diagram is the same as that given previously for $d=4$
except that each
point now represents the product of a $(d-2)$-sphere with a $p$-plane.

For appropriate values of $d$ and $p$ the metric \threel\ includes all the
known
extreme, non-dilatonic, extended object solutions of higher-dimensional
supergravity
theories. These are the membrane [\DS] and fivebrane [\Gu] solutions of
eleven-dimensional supergravity, the self-dual three-brane of ten-dimensional
supergravity [\HS] and the self-dual string of six-dimensional supergravity
[\DL].


\chapter{ Generalization to multi-$p$-branes}

By introducing the new radial coordinate $\rho$ via
$$
r^{d-3} = \rho^{d-3} + \mu^{d-3}\ ,
\eqn\ffoura
$$
and a new set of cartesian transverse space coordinates $\{ x^i;\, i=1\dots
d-1\}$ such that $\rho =\sqrt{\sum x^i x^i}$, the extreme $p$-brane solution of
\threel\ can be written in the `isotropic'
form
$$
\eqalign{
ds^2 &= H^{-{2\over p+1}}\big(-dt^2 + d{\bf y}\cdot d{\bf y}\big) + H^{2\over
d-3}\big(d{\bf x}\cdot d{\bf x}\big)\cr
{1\over (d-2)!}&\varepsilon^{ij_1\cdots j_{(d-2)}} F_{j_1\cdots j_{(d-2)}} =
\sqrt{{(d+p-2)\over 2(p+1)(d-3)}}\ \partial_i H }
\eqn\ffourb
$$
where $\varepsilon^{i_1\cdots i_{(d-1)}}$ is the constant alternating tensor
density of the Euclidean $(d-1)$-space, and
$$
H= 1+ \Big({\mu\over \rho}\Big)^{d-3}\ .
\eqn\ffourc
$$
Because of a force balance between the gravitational and Coulomb forces,
a multi-$p$-brane solution is possible in which $H$ is any solution of
Laplace's
equation in Euclidean $(d-1)$-space with $k$ point `sources' at ${\bf x}={\bf
x}_a$, i.e.
$$
H = 1 + \Big({\mu\over \rho}\Big)^{d-3} + \sum_{a=2}^k
\Big({\mu_a\over |{\bf x}-{\bf x}_a|}\Big)^{d-3}
\eqn\ffourd
$$
where we have used the translational invariance to locate the
first point `source' at the origin. The `sources' are actually the horizons of
the individual $p$-branes and if there is an analytic continuation of
the metric through these horizons then it will not be necessary to suppose that
there are material point sources there.

To investigate whether this analytic continuation is possible we expand the
terms
in the sum of \ffourd\ as an analytic series of the form
$$
\sum_{\ell=0}^\infty a_\ell \rho^\ell P_\ell
\eqn\ffoure
$$
where $P_\ell$ are a complete set of harmonics on $S^{d-2}$. One may now
attempt to
analytically continue through the horizon at $\rho=0$ as before. The
variable analogous to $\omega$ of \twom\ is given by
$$
(p+1)\mu\, \omega = \Big[1+\Big({\mu\over\rho}\Big)^{d-3}\Big]^{-{1\over p+1}}
\eqn\ffourf
$$
or
$$
\eqalign{
\rho &= \mu\big[(p+1)\mu \omega]^{p+1\over d-3}
\Big[1- \big[(p+1)\mu\, \omega]^{p+1}\Big]^{-{1\over d-3}} \cr
&= f(\omega)\omega^{p+1\over d-3} }
\eqn\ffourg
$$
where $f$ is an analytic function at $\omega=0$. We deduce that
$$
\eqalign{
H &= \Big[(p+1)\mu\, \omega\Big]^{-(p+1)} + \sum_{\ell=0}^\infty a_\ell
(\omega)\omega^{(p+1)\ell\over d-3} P_\ell \cr
&= \Big[(p+1)\mu\, \omega\Big]^{-(p+1)}\Big[ 1+ \sum_{\ell=d-3}^\infty
b_\ell(\omega,
\Omega) \omega^{(p+1)\ell\over d-3}\Big] }
\eqn\ffourh
$$
where the $a_\ell$ are analytic functions of $\omega$ and the $b_\ell$ are
analytic
functions of $\omega$ that are also functions on the $(d-2)$-sphere. The
leading term is similar to what we had before, and corresponds to a regular
horizon at $\om=0$. The higher order terms involve powers of $\om^\nu$ where
$$
\nu \equiv {p+1\over d-3}
\eqn\ffouri
$$
The $p$-brane
solution \ffourb\ will therefore have an analytic continuation through the
horizon at $\rho=0$ (and then, evidently, through any of the $k$ horizons)
provided
that $\nu$
is an integer.

This condition is always satisfied for $d=4$. In this case, since $\nu$ is an
even
integer when $p$ is odd,  we recover the same criterion as before for
the symmetric
extension through the horizon, and consequent complete non-singularity. When
$d>4$
the condition \ffouri\ is harder to satisfy. The $\nu=1$ case includes the
$D=6$ self-dual string and the $D=10$ self-dual three-brane. Both of these
cases have $p$ odd and hence are completely nonsingular as a single object,
but when more than one object is present, the extension
through the horizon is now asymmetric and there are likely to be singularities
in
the interior regions. This is perhaps an indication that the nonsingularity of
the single-core solutions is not a stable feature.  The most interesting
example of
a $\nu=2$ case for $d>4$ would appear to be the multi-version of the $D=11$
fivebrane, which is a solution of 11-dimensional supergravity that interpolates
between eleven-dimensional Minkowski spacetime and the $S^4$ compactification
to
$adS_7$ [\GT]. An interesting point about this case is that if one chooses a
solution of the Laplacian for which $H\rightarrow 0$ as $\rho\rightarrow
\infty$,
i.e.
$$
H = \sum_{a=1}^k
\Big({\mu_a\over |{\bf x}-{\bf x}_a|}\Big)^3\ ,
\eqn\ffourd
$$
the resulting non-singular spacetime can be viewed as one that interpolates
between
many different, effectively seven-dimensional, spacetimes with differing values
(determined by the constants $\mu_a$) of the cosmological constant. This is a
`local
compactification' of the type envisaged by van Baal et al. [\vB].

If $\nu$ is not an integer, then we have not found coordinates in which
the metric is smooth at the horizon. While it is still possible that such
coordinates exist, it is more likely that the horizons are only $C^k$
for a finite $k$.
This is similar to what was found for multi-black holes in de Sitter space
[\BHKT]. The
simplest example of noninteger $\nu$ is multi-black holes
in five dimensions (coupled to a
three form) which has
$\nu = 1/2$. The smoothness of these solutions deserves to be investigated
further.


\chapter{ An energy bound for $p$-branes}

A sufficient condition for the stability, subject to given boundary
conditions, of a solution to a classical field theory is that its energy
saturate a
lower bound on the energy of all field configurations satisfying the boundary
conditions. As mentioned in the introduction, certain `extreme' charged black
hole solutions are known to saturate such a bound. Here we consider the
circumstances under which a similar bound can be derived for the more general
case of a $p$-brane solution of a $D$-dimensional field theory. The solutions
whose stability we wish principally to consider are infinite planar $p$-branes,
for
which the total energy is clearly infinite. In this case, however, the
relevant concept is the total energy per unit $p$-volume and for these purposes
we
can suppose that the $D$-dimensional spacetime has the topology ${\bf
R}^d\times T^p$
and that the $p$-brane is wrapped around the $p$-torus. The energy per unit
$p$-volume is then the (now finite) total energy divided by the volume of the
$p$-torus. Note that the concept of spatial infinity is now replaced by
`transverse
spatial infinity' and the $(D-2)$-sphere at spatial infinity is replaced by the
$(D-2)$-dimensional surface $S^{d-2}\times T^p$.

In the gravitational context, to define the total energy
per unit $p$-volume of a classical
$p$-brane solution one needs a vector field that is asymptotically timelike
and Killing near transverse spatial infinity. As in the
$p=0$ case [\Wi], this vector field can be replaced by a complex commuting
spinor
field $\epsilon$ (of the $D$-dimensional Poincare group) that is asymptotic to
a
constant spinor $\epsilon_\infty$. The total, transverse, $d$-momentum per-unit
$p$-volume, $P_\mu$, can then be expressed via the integral
$$
\bar\epsilon_\infty (\Gamma^\mu P_\mu)\epsilon_\infty = {1\over2
V_p\Omega_{(d-2)}}\oint
\!dS_{mn} E^{mn}\ ,
\eqn\foura
$$
where $dS_{mn}$ is the surface element of the $(D-2)$-surface $S^{d-2}\times
{\bf
T}^p$ at transverse spatial infinity, $V_p$ is the volume of the $p$-torus,
$\Omega_{(d-2)}$ is the volume of the unit $(d-2)$-sphere, and
$$
E^{mn} ={1\over2}\bar\epsilon\Gamma^{mnp}\nabla_p\epsilon + c.c.\
\eqn\fourb
$$
is the $D$-dimensional Nester tensor.
The conjugate spinor $\bar\epsilon$ is defined by
$$
\bar\epsilon \equiv \epsilon^\dagger \Gamma^{\underline 0}
\eqn\fourc
$$
where underlining indicates an orthonormal frame index. Note that
$$
\Gamma^{\underline 0}(\Gamma^{(k)})^\dagger \Gamma^{\underline 0}
= -(-1)^{k(k+1)\over2} \Gamma^{(k)}
\eqn\fourd
$$
where $\Gamma^{(k)}$ is an antisymmetrized product of $k$ gamma matrices.

We wish to consider all metrics on $R^d \times T^p$ that are asymptotically
flat in the sense that the deviation $h_{mn}$
of the metric from flat space
falls off asymptotically like $O(1/r^{d-3})$ where $r$ is a transverse
radial distance. Letting $x^\mu$ denote the
coordinates on $R^d$ and $y^i$ denote the coordinates on $T^d$,
we further require that $h_{\mu i}$ and
the derivative of $h_{mn}$ with respect to $y^i$ both fall off like
$O(1/r^{d-4})$. This ensures that there are no Kaluza-Klein type charges
in the $d$ dimensional space, and that
the leading order deviation is independent of $y^i$.
Under these conditions the integral \foura\ can be written as
$$
\bar\epsilon_\infty (\Gamma^\mu P_\mu)\epsilon_\infty = {1\over
2\Omega_{(d-2)}}
\oint
\!dS_{\mu\nu} E^{\mu\nu}\ ,
\eqn\foure
$$
where the integral is now over the $(d-2)$-sphere at infinity. One can show
that
$$
M = \sqrt{- P^\mu P_\mu}\ .
\eqn\fourn
$$
equals the standard ADM mass, in units for which $G=1$, of a spacetime which is
asymptotically $R^d \times T^p$ [\BKK]. For the $p$-brane metrics of
\threel\ one finds that
$$
M= {(D-2)\over 2(p+1)} \mu^{d-3}\ .
\eqn\ADMmass
$$
so that the mass to charge ratio of these solutions is
$$
{M\over |Q|} = \sqrt{ (D-2)\over 2(p+1)(d-3)}
\eqn\ratio
$$

We now have the necessary ingredients for a proof of the positivity of
the energy per unit $p$-volume but we are interested in obtaining a stronger
lower bound in terms of the electric or magnetic charge per unit $p$-volume. To
this end we now introduce an $n$-form abelian field-strength,
$n$ to be specified shortly,
and the associated modified Nester tensor
$$
\hat E^{mn} = {1\over2}\bar\epsilon\Gamma^{mnp}\hat
\nabla_p\epsilon + c.c.\ ,
\eqn\fourg
$$
via the modified covariant derivative
$$
\hat\nabla_p = \nabla_p + {c\over (D-2) n!}\big[ (n-1)
\Gamma_p{}^{m_1\dots m_n} -n(D-n-1)\delta_p^{m_1}
\Gamma^{m_2\dots m_n}\Big]F_{m_1\dots m_n}
\eqn\fourh
$$
where $c$ is a constant. Note that there are two possible terms proportional to
the
components of the $n$-form $F$. The relative coefficient is fixed by
requiring the cancellation of the `$(\epsilon\Gamma\nabla\epsilon)F$' terms in
the
calculation to follow\foot{  It is also required by considerations of
linearized
supersymmetry [\TvN], but we do not wish to {\it assume} supersymmetry here.}.
This
cancellation also requires that
$$
\bar c = (-1)^{n(n+1)\over2} c\ .
\eqn\fouri
$$
where $\bar c$ is the complex conjugate of $c$; thus $c$ is either real or
purely
imaginary. A short exercise in gamma matrix algebra yields
$$
\hat E^{\mu\nu}=E^{\mu\nu} - {c\over n!}
(\bar\epsilon\Gamma^{\mu\nu p_1\dots p_n}\epsilon) F_{p_1\dots p_n}
-{c\over (n-2)!}
(\bar\epsilon\Gamma^{p_3\dots p_n}\epsilon)
F^{\mu\nu}{}_{p_3\dots p_n}
\eqn\fourj
$$

For definiteness we shall now suppose that the $p$-brane is purely magnetic, in
which case the last term vanishes and $n=d-2$. We shall also assume that $d\ge
4$, so $n\ge 2$. Thus, in what follows the integers
$d$ and $n$ are related to the spacetime dimension $D$ and the spatial
dimension
$p$ of the extended object by
$$
\eqalign{
d &= D-p \ge 4 \cr
n &= d-2 = D-p-2 \ .}
\eqn\afourj
$$
We further suppose, as part of the boundary conditions to be satisfied by the
fields,
that the only components of $F$ that contribute to the integral over the
surface at
infinity are $F_{\mu_1\dots
\mu_n}$. Then
$$
{1\over2 V_p\Omega_{(d-2)}}\oint \!dS_{mn} \hat E^{mn}= \bar\epsilon_\infty [
\Gamma^\mu P_\mu - c Q \Gamma_*]\epsilon_\infty
\eqn\fourk
$$
where
$$
Q = {1\over2(d-2)!\Omega_{(d-2)}} \oint
dS_{\mu\nu}\varepsilon^{\mu\nu\rho_3\dots\rho_d}F_{\rho_3\dots\rho_d}
\eqn\fourl
$$
is the magnetic charge per unit $p$-volume and
$$
\Gamma_* = \Gamma^{\underline 0}\Gamma^{\underline 1}\cdots\Gamma^{\underline
{(d-1)}}\ .
\eqn\fourm
$$
If we can show that the left hand side of \fourk\ is positive semi-definite
then,
using \fouri\ , we can derive the bound
$$
M\ge |c||Q|
\eqn\afourm
$$
on the total mass per unit $p$-volume.

Assuming that any horizons or singularities are
the result of gravitational collapse from a configuration without horizons or
singularities, we may choose a spacelike hypersurface, on which all fields are
regular, such that its only boundary is the surface at transverse spatial
infinity. Then, Gauss' law can be used to write the surface integral on the
left
hand side of \fourk\ as a volume integral:
$$
{1\over 2\Omega_{(d-2)}}\oint dS_{mn}\hat E^{mn} = -{1\over \Omega_{(d-2)}}\int
dS_n
\nabla_m \hat E^{mn}
\eqn\fouro
$$
Then, using the field equation
$$
G_{mn} = 2T_{mn}(F)\ ,
\eqn\fourp
$$
where
$$
T_{mn}(F) \equiv {1\over (n-1)!}( F_{mr_1\dots r_{(n-1)}} F_{n}{}^{r_1\dots
r_{(n-1)}} -{1\over 2n} g_{mn} F^2) \ ,
\eqn\fourq
$$
one finds after a long calculation that
$$
\eqalign{
\nabla_m \hat E^{mn} = &\overline{\hat\nabla_m\epsilon}\, \Gamma^{mnp}
\hat\nabla_p\epsilon - {c\over (n-2)!}(\bar\epsilon \Gamma_{p_3\dots
p_n}\epsilon)
\nabla_m F^{mnp_3\dots p_n} \cr
&- \Big(1- {2(D-n-1)(n-1)|c|^2\over (D-2)}\Big)(\bar\epsilon\Gamma_m\epsilon)\,
T^{mn}(F) \cr
& -{2|c|^2\over (D-2)}\sum_{k=1}^\infty
{(-1)^k\big(n^2 -nD + (2k+1)D -4k-1\big)\over (n-2k-1)! (2k)!(2k+1)!}\times\cr
& \times (\bar\epsilon\Gamma_{m p_1\dots p_{2k}q_1\dots q_{2k}}\epsilon)\,
T^{mn p_1\dots p_{2k}q_1\dots q_{2k}}(F)\ ,}
\eqn\fourr
$$
where
$$
\eqalign {
T^{mn p_1\dots p_{2k}q_1\dots q_{2k}}(F) &\equiv
\big[ F^{mp_1\dots p_{2k}r_1\dots r_{(n-2k-1)}} F^{n q_1\dots
q_{2k}}{}_{r_1\dots
r_{(n-2k-1)}} \cr
&\qquad -{(2k+1)\over 2(n-2k)}g^{mn}
F^{p_1\dots p_{2k}r_1\dots r_{(n-2k)}}F^{q_1\dots q_{2k}}{}_{r_1\dots
r_{(n-2k)}}\big]\ . }
\eqn\afourr
$$
Note that the potentially infinite sum, which contains only terms of the
form `$\bar\epsilon\Gamma^{(4k+1)}\epsilon F^2$', truncates itself
as soon as $4k+1$ exceeds either $D$ or $2n-1$.
We want all terms on the right hand side of \fourr, except the first, to cancel
on
using the $F$ field equation. In this case we will have
$$
{1\over 2\Omega_{(d-2)}}\oint dS_{mn}\hat E^{mn}  = -{1\over
\Omega_{(d-2)}}\int\!
dS_n \overline{\hat\nabla_m\epsilon}\, \Gamma^{mnp} \hat\nabla_p\epsilon
\eqn\fours
$$
which can be shown to be positive semi-definite by choosing $\epsilon$
to satisfy the usual modified Witten condition.

At this point we need the field equation for $F$ in order to proceed. The field
equation \fourp\ is derivable from an action of the form
$$
S= \int d^D x \sqrt{-g}\big(R - {2\over n!}F_n^2 + {\rm possible}\
\varepsilon AFF\ {\rm term} \big)\ ,
\eqn\fourt
$$
where we allow for a possible abelian Chern-Simons (CS) term in the action when
$D=3n-1$ because it does not affect the stress tensor $T_{mn}(F)$.
The simplest case for which this term may appear is $D=5$ and $n=2$.

We now turn to the investigation of when the necessary cancellations can occur.
The
simplest possibility occurs when all the $\bar\epsilon\Gamma^{(4k+1)}\epsilon
F^2$
terms vanish for $k\ge1$. Given that $d\ge 4$, this happens only for $D=4$ and
$n=2$, i.e. $p=0$. In this case no CS term is possible so the $F$ field
equation is
simply
$$
\nabla_m F^{mn}=0\ .
\eqn\fourv
$$
By choosing
$$
|c|=1
\eqn\fouru
$$
we ensure the cancellation of the remaining $(\bar\epsilon\Gamma_m\epsilon)
T^{mn}(F)$ terms and \fourr\ now reduces to \fours\ as required. We thus
recover the bound $M\ge|Q|$ [\GH] on magnetically-charged particle-like
solutions of
four-dimensional Maxwell-Einstein theory.

For $D>4$ the derivation of a similar bound is necessarily more
complicated because a $\bar\epsilon\Gamma^{(5)}\epsilon F^2$ term, at least,
then
appears on the right hand side of \fourr. One might hope that its coefficient
could vanish but this happens only if
$$
n= {1\over2}(D \pm \sqrt{(D-2)(D-10)})\ ,
\eqn\fourw
$$
which is never satisfied for $2<D<9$. In fact the only real integer solutions
of
this equation, with $n\ge2$, that we have found are
\vskip 0.3cm
(i) $D=10$ and $n=5$\hskip 1cm (ii) $D=11$ and $n=4$ or $n=7$
\vskip 0.3cm
The absence of the $\bar\epsilon\Gamma^{(5)}\epsilon F^2$ term in these cases
will be seen to be of great importance for the derivation of a Bogomol'nyi-type
bound for the
ten-dimensional self-dual threebrane and the eleven-dimensional fivebrane
and membrane. In all these cases, however, and whenever $n\ge4$, the
$\bar\epsilon\Gamma^{(9)}\epsilon F^2$ term also appears on the right hand side
of
\fourr, unless its coefficient vanishes, which happens only if
$$
n= {1\over2}(D \pm \sqrt{(D-2)(D-18)})\ .
\eqn\fourx
$$
This cannot be satisfied if $2<D<18$. Thus, for $D>4$ we always have, {\it at
least}, either a $\bar\epsilon\Gamma^{(5)}\epsilon F^2$ or a
$\bar\epsilon\Gamma^{(9)}\epsilon F^2$ term on the right hand side of \fourr.
In
order to derive a Bogomol'nyi-type
bound we must find a way to cancel these terms. The
freedom that we have to achieve this cancellation consists of, when applicable,
(i)
the possibility of choosing $F$ to be self-dual and spinor $\epsilon$ to be
chiral,
(ii) the choice of the constant $|c|$, and (iii) the choice of coefficient of
the CS
term in the action. Consideration of these possibilities leads to two
mechanisms for
removal of the unwanted terms in \fourr, which we shall now examine in detail.
\vskip 0.3cm
(a) If $D=4k'+2$ the final $k=k'$ term $\bar\epsilon\Gamma^{(4k'+1)}\epsilon
F^2$
term in the sum of \fourr\ is equivalent to a
$(\bar\epsilon\Gamma^{(1)}\epsilon)\,
\varepsilon F^2$ term, where $\varepsilon$ indicates a Levi-Civita tensor, if
$\epsilon$ is chosen to be a chiral spinor. We can get then rid of the
$\varepsilon$ by choosing $n=2k'+1$ and declaring $F$ to be self-dual (note
that this
is possible precisely for dimensions $D=4k'+2$). Remarkably, this term is then
of the
form $\bar\epsilon \Gamma_m\epsilon\, T^{mn}(F)$ so it can be combined with the
other stress tensor terms on the right hand side of \fourr. These terms then
cancel if $|c|$ is chosen such that
$$
|c|^2 = {1\over 2(n-1)} = {1\over 2(p+1)}\ .
\eqn\afourx
$$
By this means we can cancel
the ($k=1$) $\bar\epsilon\Gamma^{(5)}\epsilon F^2$ term for $D=6$, $n=3$, and
the ($k=2$)
$\bar\epsilon\Gamma^{(9)}\epsilon F^2$ term for $D=10$, $n=5$. Recalling that
the
coefficient of the $\bar\epsilon\Gamma^{(5)}\epsilon F^2$ term cancels when
$D=10$
and $n=5$ we see that the bound
$$
M\ge {1\over \sqrt{2(p+1)}}|Q|\ .
\eqn\bfourx
$$
can be established for the self-dual $D=6$ string ($p=1$) and $D=10$ threebrane
($p=3$). Observe that the square of the mass to charge ratio of a solution
saturating this bound is half the value, given in \ratio, for the extreme
$p$-brane solution of \threel. This is because the solution
\threel\ describes a magnetically charged $p$-brane. The self-dual solution
can be obtained simply by performing a duality rotation on $F$. This does not
change the metric, but reduces the charge (as defined in \fourl) by $\sqrt 2$.
These solutions therefore saturate the bound.

For $k'=3$, i.e. $D=14$ we could cancel the $k=3$,
$\bar\epsilon\Gamma^{(13)}\epsilon
F^2$, term by this method but we would still be left with the
$\bar\epsilon\Gamma^{(5)}\epsilon F^2$ and $\bar\epsilon\Gamma^{(9)}\epsilon
F^2$
terms, neither of which has a vanishing coefficient. They might cancel
each other, however, since $5+9=14$ and $F$ is self-dual. Similar cancellations
might
occur for $k'>3$. In view of the fact that the cancellations possible for
$k'=1$ and
$k'=2$ are realized by supergravity theories, a feature that is also shared by
the
alternative mechanism to be explained below, and the fact that there are no
supergravity theories with $D>11$ we suspect that the required cancellations
for
e.g. $D=14$ do not actually occur. We have not undertaken the calculations
necessary
to verify this because we are not aware of any interest in e.g. self-dual
fivebranes
in $D=14$.

\vskip 0.5cm
(b)
When $D\ne 4k'+2$ we must choose
$$
|c|^2 = {D-2\over 2(n-1)(D-n-1)}
\eqn\cfourx
$$
or equivalently,
$$
|c|^2 = {D-2\over 2(p+1)(d-3)}
\eqn\fourxx
$$
in order to cancel the stress tensor terms in \fourr, since they cannot cancel
with
other terms in the sum. For $D>4$ we must therefore find some other means of
cancelling the non-zero terms in the sum over $k$ in \fourr. When $D=3n-1$ we
have
the possibility of a Chern-Simons term in the action. Use of the $F$ field
equation
then produces a term of the form $(\bar\epsilon\Gamma^{(n-2)}\epsilon)\,
\varepsilon
F^2$ so we have a possibility of cancelling the
$(\bar\epsilon\Gamma^{(4k'+1)}\epsilon) F^2$ term when $n=2k'$, i.e $D=3k'-1$,
and $k=k'$. For $k'=1$ this implies $D=5$ and $n=2$, and since there are no
$\bar\epsilon\Gamma^{(9)}\epsilon F^2$, or higher, terms in this case the
required
cancellation is complete and we find the bound
$$
M\ge {\sqrt{3}\over 2} |Q| \qquad (D=5, p=1)
\eqn\dfourx
$$
for string-like solutions of five-dimensional supergravity (which includes the
needed
CS term with just the right coefficient). The same formula was found for
electric-type particle-like solutions of five-dimensional supergravity
[\GKLTT],
which is presumably a reflection of the fact that for $D=5$ a magnetic-type
string
is dual to an electric-type particle.

For $k'=2$ this mechanism allows the cancellation of the
$\bar\epsilon\Gamma^{(9)}\epsilon F^2$ term when $D=11$ and $n=4$. Since this
is
precisely one of the cases for which the coefficient of the
$\bar\epsilon\Gamma^{(5)}\epsilon F^2$ vanishes the required cancellation is
again
complete and we establish the bound
$$
M\ge {1\over 2} |Q| \qquad (D=11,p=5)
\eqn\efourx
$$
on the mass per unit area of a fivebrane. Both of the bounds \dfourx\ and
\efourx\ are saturated by the respective special cases of the extreme
$p$-brane solution of \threel, as can be seen by comparing \ratio\ and
\fourxx.

For $k'>2$ the inclusion of a CS term cannot cancel either the
$\bar\epsilon\Gamma^{(5)}\epsilon F^2$ or the $\bar\epsilon\Gamma^{(9)}\epsilon
F^2$ term, at least one of which must have a non-vanishing coeficient. Thus
$D=5$
supergravity and $D=11$ supergravity are the only cases for which the required
cancellations can be achieved by the inclusion of the CS term.
\vskip 0.5cm

The conclusion is that, for $D\ge4$ and $n\ge 2$, a Bogomol'nyi bound
exists for magnetically charged {\it non-dilatonic} $p$-brane solitons for the
following cases (and only for these cases if self-dual fivebranes in $D=14$ and
their generalizations to higher dimensions are excluded):
\vskip 0.3cm
(i) $D=4$ and $p=0$
\vskip 0.3cm
(ii) $D=5$ and $p=1$ with CS term
\vskip 0.3cm
(iii) $D=6$ and $p=1$ with self-dual $F$
\vskip 0.3cm
(iv) $D=10$ and $p=3$ with self-dual $F$
\vskip 0.3cm
(v) $D=11$ and $p=5$ with CS term.
\vskip 0.3cm
Each of these cases is realized by a supergravity theory. We expect that a
similar
analysis for electric-type $p$-branes would lead to the same conclusion for the
electric duals of the above cases. This would allow us to extend the above list
of
$(D,p)$ values to include
\vskip 0.5cm
(vi) $D=5$ and $p=0$ with CS term
\vskip 0.3cm
(vii) $D=11$ and $p=2$ with CS term
\vskip 0.3cm
\noindent Case (vi) is realized by black
hole solutions of D=5 supergravity [\GKLTT] and case
(vii) by membrane solutions of $D=11$ supergravity [\DS].

For theories with a Bogomol'nyi-type bound, configurations which
saturate this bound must be stable. They must also admit Killing spinors,
i.e. solutions of
$$
\hat\nabla_m \epsilon =0
\eqn\foury
$$
for non-zero $\epsilon$. This is because if $M=|c| |Q|$, one
can choose $\epsilon_\infty$ so that
$$
[\Gamma^\mu P_\mu - c\Gamma_* Q ]\epsilon_\infty =0 \ .
\eqn\fourz
$$
It then follows from \fourk\ and \fours\ that there exists an $\epsilon$
which approaches $\epsilon_\infty$ asymptotically and
satisfies $\hat\nabla_m \epsilon =0$. In fact, for the multi-$p$-brane
solution of \ffourb\ one can show that the Killing spinor satisfying
$$
{c\over |c|}\Gamma^{\underline{0}}\Gamma_* \epsilon = \pm \epsilon
$$
is given by
$$
\epsilon = H^{\pm {1\over 2(p+1)}}\epsilon_\infty
\eqn\Kilspin
$$


\chapter{Discussion}

We have shown that the singularities in a number of extreme dilatonic
black holes and $p$-branes can be resolved by viewing the solutions
as reductions of higher dimensional  objects. This raises a few
issues which we address in this section. The first concerns the
relation between our nonsingular solutions with horizons \twol\ and \threel,
whose maximal analytic extensions
are represented by the Carter-Penrose diagram of Fig. 1, and the
singularity theorems. There is nothing special about four dimensions
in the proof of the singularity theorems. Higher dimensional solutions
which satisfy the conditions of the theorems must be geodesically incomplete.
The most important condition in our case is the existence of a compact
trapped surface. Our nonsingular $p$-branes simply do not have any
surfaces of this type. The horizon is only a marginally trapped surface, since
the null generators have zero convergence, and it is not compact if the
extra dimensions are not periodically
identified. If they are identified, then the solution
is geodesically incomplete anyway.

It is instructive to consider the relation between our
non-singular metrics and the singularity
theorems in more detail. It will be sufficient to consider the simplest case
of the
extreme black string in a five-dimensional spacetime. The singularity
theorems consist of a global part (e.g. the existance of a trapped surface)
and a local part which concerns the focussing of geodesics.
Consider a small bundle of
null rays each moving radially inwards ($\theta$ and $\phi$ constant) with
constant $y=x^5$ coordinate; these rays are represented by a straight line at
45${}^\circ$ in the CP diagram. As they pass from one asymptotically flat
region
to another their cross-section decreases and then increases. Since the
five-dimensional energy-momentum tensor satisfies the null energy condition,
Raychaudhuri's equation tells us that the null rays must encounter a caustic.
The area of a two-sphere in the $\theta$ and $\phi$ directions is proportional
to $r^2$ and since $r$ is bounded below by $\mu$ there is no caustic in these
directions. Since these geodesics have $x^5$ constant, they are also
null geodesics in the four dimensional space. It follows that the
{\it four-dimensional} metric must violate the
null energy condition or be singular, according to the choice of `conformal
gauge'; in the five-dimensional Einstein conformal gauge the null energy
condition is violated while the metric in the four-dimensional Einstein
conformal
gauge is singular. However, the cross section in the fifth, $x^5$, direction is
proportional to $\Big(1-{\mu\over r}\Big)$ and this {\it does} focus at
$r=\mu$. In other words, the rays do encounter a caustic but it is by
approaching each other in the $x^5$ direction. Of course, the fact that the
rays encounter a caustic does not mean that there is a singularity (this is
where the global condition is needed). In fact,
there is no singularity. However, this example does serve to illustrate the
fact that when considering focussing (which is an important part
of the
singularity theorems) one must consider not just the four physical spacetime
dimensions but the extra dimensions too.

We next consider how general
the method we have described is for resolving singularities.
Is it possible that all physical singularities in four dimensions
could be resolved in this way? Unfortunately, the answer appears to be
no, at least
not without serious modification. The point is that in our approach the
higher dimensional metric includes the lower
dimensional one rescaled by a power of the dilaton. So roughly speaking,
only singularities in the Ricci tensor and not the Weyl tensor
can be removed this way.\foot{This is not strictly true since the dilaton
is also diverging, and a singular conformal factor can transform a
diverging $C^2$ into one that remains finite.}
For example, it seems unlikely that
one can find a  regular higher dimensional solution whose reduction
yields the Schwarzschild metric.

A related point is that
given a theory with a dilaton, there is some ambiguity about which
metric one should consider in deciding  whether  a solution is singular.
One often considers the metric with the standard Einstein action, but other
metrics related by a conformal rescaling
may be physically important depending on the coupling to matter.
For example, if one rescales the extreme dilaton black hole metric \twob\
by $e^{2\phi/a}$, the resulting metric is geodesically complete. So if one
adds matter which couples to this metric, the solution might be called
nonsingular even in four dimensions. (This ambiguity is not present in
the higher dimensional theories we consider here which do not include a
dilaton.)

However, it should be emphasized that
we have explored only the simplest possibilities of obtaining
lower dimensional solutions from higher dimensional ones. We started with
just an Einstein-Maxwell type action in higher dimensions and did not
include
off diagonal terms in the higher dimensional metric.  It is possible that
a  more general procedure will be able to resolve other
familiar four dimensional singularities.

\vskip 1cm
\centerline{\bf Acknowledgements}
G.H. was supported in part by NSF grant PHY-9008502.

\refout

\bye